\documentclass[sn-mathphys,Numbered]{sn-jnl}%

\usepackage{graphicx}%
\usepackage{multirow}%
\usepackage{amsmath,amssymb,amsfonts}%
\usepackage{amsthm}%
\usepackage{mathrsfs}%
\usepackage[title]{appendix}%
\usepackage{xcolor}%
\usepackage{textcomp}%
\usepackage{subcaption}
\usepackage{manyfoot}%
\usepackage{booktabs}%
\usepackage{algorithm}%
\usepackage{algorithmicx}%
\usepackage{algpseudocode}%
\usepackage{listings}%

\theoremstyle{thmstyleone}%
\theoremstyle{thmstyletwo}%

\theoremstyle{thmstylethree}%

\begin{document}

\title[Article Title]{A novel application of Shapley values for large multidimensional time-series data: Applying explainable AI to a DNA profile classification neural network}

\author[1]{\fnm{Lauren} \sur{Elborough}}\email{lauren.elborough@adelaide.edu.au}

\author[2,3]{\fnm{Duncan} \sur{Taylor}}\email{duncan.taylor@sa.gov.au}

\author*[1]{\fnm{Melissa} \sur{Humphries}}\email{melissa.humphries@adelaide.edu.au}

\affil[1]{\orgdiv{Adelaide Data Science Centre, School of Computer and Mathematical Sciences}, \orgname{University of Adelaide}}

\affil[2]{\orgname{Forensic Science SA, South Australia 5001}}

\affil[3]{\orgdiv{College of Science and Engineering}, \orgname{Flinders University}}


\abstract{The application of Shapley values to high-dimensional, time-series-like data is computationally challenging - and sometimes impossible.  For $N$ inputs the problem is $2^N$ hard. In image processing, clusters of pixels, referred to as superpixels, are used to streamline computations. This research presents an efficient solution for time-seres-like data that adapts the idea of superpixels for Shapley value computation. Motivated by a forensic DNA classification example, the method is applied to multivariate time-series-like data whose features have been classified by a convolutional neural network (CNN). In DNA processing, it is important to identify alleles from the background noise created by DNA extraction and processing. A single DNA profile has $31,200$ scan points to classify, and the classification decisions 
must be defensible in a court of law. This means that classification is routinely performed by human readers - a monumental and time consuming process. The application of a CNN with fast computation of meaningful Shapley values provides a potential alternative to the classification. This research demonstrates the realistic, accurate and fast computation of Shapley values for this massive task.}


\maketitle

\section{Introduction}\label{sec1}

This research provides a workable solution for applying Shapley values to high-dimensional time-series style problems. This approach provides a general solution demonstrated through the application to DNA profile classification. 

The black box limitation of AI models is particularly of interest when there is an impact on industries where a high level of trust is needed, for example in medical settings, finance, autonomous vehicles, court, and forensics \cite{barredo_arrieta_explainable_2020, chou_counterfactuals_2022}. In forensic science explainability is of particular importance for ensuring the fair admission and cross-examination of any new evidence evaluation technique.  Explainable artificial intelligence (XAI) techniques are a suite of techniques that seek to improve the interpretability of AI models to understand how they are making classifications and reaching decisions \cite{barredo_arrieta_explainable_2020}. 

Shapley values are a common choice for XAI applications because they describe the contributions of different inputs to the ultimate classification or decision. First presented by Lloyd S. Shapley in 1953, for use in in cooperative game theory, the motivating aim of Shapley values was to assess the promise of each player in a multiperson game \cite{shapley_games_1952}. The technique identifies ``variables of importance" and, as such, has found extensions in a variety of settings including economics, business, politics, machine learning, and online marketing \cite{hart_shapley_2017}. Classified as an XAI technique, Shapley values allocate credit for a model’s output to each of the input features, hence improving the understandability of the model. However, there are limitations to the use of Shapley values, one of which relates to the size of the problem being studied. 

As the number of inputs of an AI problem increases (such as in image classification), calculation of Shapley values becomes infeasible. For N inputs the problem is $2^N$ hard. One method used to overcome this problem in image processing, is to break the inputs into clusters of pixels, referred to as superpixels \cite{dardouillet_explainability_nodate}. For time-series style data, what counts as a ``superpixel" is not clear. Treating the series as an image does not lead to a practically solvable problem due to either having too many superpixels to consider, or requiring superpixels that are so large that they lack the resolution to meaningfully inform the user as to the decision-making logic.  

The superpixel limitation is addressed in our work through targeted partitioning and isolation of the data. We show the successful development of a system which calculates Shapley values, through the use of the popular extension on Shapley values, Kernel SHAP \cite{lundberg_unified_2017}. This research presents an extension to the image-based superpixel framework, demonstrating the suitability for vectorized data types with time-series-like dependence. 

The application of this approach is in a forensic statistical context, for analysing DNA profile data, both confirms the potential for improving insights into automated DNA profile classification, and presents a computationally feasible possibility for the application of Shapley values to similar data types.  

\subsection{Motivating example: DNA profiling}

\begin{figure}[h!]
    \centering
    \includegraphics[width = \textwidth]{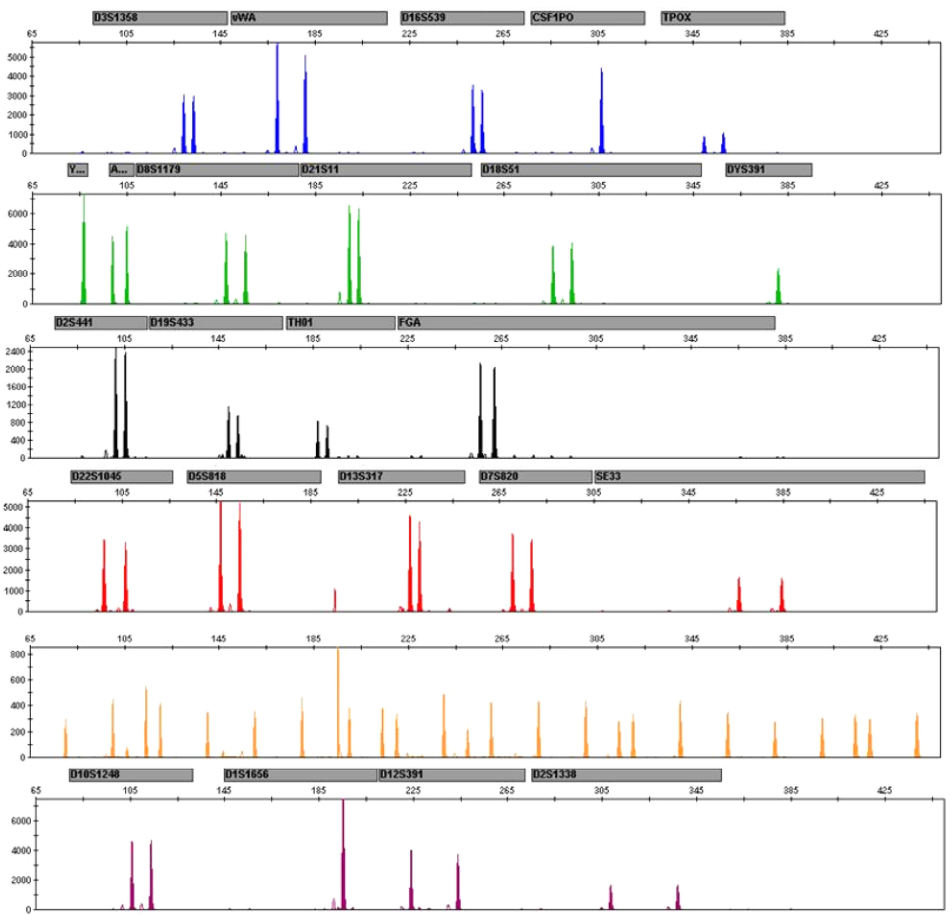}
    \caption{An example of a DNA profile for the GlobalFiler$^{TM}$ profiling system. The profile has six dye lanes, and measures the relative fluorescence units (RFUs) on the y-axis are measured over a sequence of base pairs on the x-axis. The profile contains both DNA and artefacts, to be removed during the reading. The boxes above each dye lane represent different 24 different regions of DNA that are targetted in GlobalFiler$^{TM}$ during the PCR process.}
    \label{fig:dna}
\end{figure}

An important component of many criminal investigations is DNA profiling. In forensic DNA laboratories, DNA samples are processed to produce electropherograms (EPGs), commonly referred to as DNA profiles. The pathway from sample to EPG is complex (see \cite{fund_forens} for a detailed summary) in which the fluorescent signals are separated based on their output wavelength, leading to  multiple time-series-style outputs of fluorescently tagged synthetic DNA fragments. Different commercially available DNA profiling kits utilise different combinations and numbers of fluorophores (often called dyes). We demonstrate the application of our XAI technique to data produced by the GlobalFiler system (produced by Thermofisher) which uses six dyes. Figure \ref{fig:dna} shows the visual representation of the EPG for one DNA profile, split across six fluorescence `dye lanes'. The y-axis in Figure \ref{fig:dna} measures the intensity of the peaks (relative fluorescent units, RFUs) across 9,000 scan points. Of these 9000 scan points, approximately 5000 are of interest, with the scan points at the beginning being in an area known as the `primer flare' and the scan points at the end being beyond the size of any targeted DNA fragments. In the process described below 5200 scan points are utilised, which are the 5000 of interest and 100 scan points on either side of padding. Other features of importance include the size (in basepairs, compared to an internal size standard) and a designation that relates to the underlying sequence. 


Since these steps are biological, there are many areas of stochastic variation in the quality of the DNA sample, efficiency of the process, instrument noise and artefacts created during the profiling process (Table \ref{tab:artefact}). As such, the peaks observed in Figure \ref{fig:dna} may not all be ``real" DNA features and the profile must be read to remove artifactual features and leave only information of interest. 

\begin{table}[]
    \centering
\begin{tabular}{c p{9cm}}
    Feature  & Description  \\
   \hline\\
    Allele & The DNA signal of interest. A gene, or pair of genes, at a specific location of a chromosome. These appear as peak(s) in the EPG and by using information about the scan point at which the peak was detected, an internal sizing standard, and an allelic ladder they are given an allelic designation that represents the number of repeating units of DNA in the underlying sequence. \\
    \\
    Baseline & The spans of data where there is no clear signal. This is representative of a near-stationary time-series of white noise.\\
   \\
    Artefact & Description \\
    \hline\\
    Stutter & Peaks that occur in the \textit{same} dye lane as an allele. These originate from errors during the replication process. Depending on the type of error a stutter peak can occur an entire repeat unit before the allele (backward stutter), an entire repeat unit after the allele (forward stutter), or half a repeat unit before the allele (half stutter).\\
    \\
    Pull-up & Peaks that occur at the same location as an allele in a \textit{different} dye lane to the true allele.\\
    \hline
    \hline
\end{tabular}
    \caption{Features and artefacts observed in DNA profiles. The goal of classification is to discern alleles from artefacts.}
    \label{tab:artefact}
\end{table}

The process of profile reading is typically completed manually by two independent human readers. Given the complexity of the information produced by modern DNA profiling systems, this can be a time-consuming undertaking \cite{taylor_teaching_2016}. After reading the profile, the readers then compare their results, resolve any discrepancies and a binary decision is made for the classification of each peak.  

In 2021, Forensic Science SA (FSSA) replaced one person in the profile reading workflow with a DNA-profile-reading convolutional neural network (CNN) \cite{taylor_using_2022}. The CNN reads in the raw fluorescent signal, for each dye lane, at each scan point and outputs classifications for each scan point in the profile into either allele (the DNA signal of interest) or one of a series of artefacts created during the profile production. While the process has been shown to work at high levels of accuracy \cite{taylor_using_2022}, a limitation to the process is that the CNN is acting as a black box, i.e. it is not clear what information in the profile is being used to designate a classification, or how the CNN is deciding on a classification between categories. This is a clear candidate for XAI application: Understanding the contributing features of the profile that lead to a classification decision may provide the depth of information required for fair admission and cross-examination in court. 

However, the size of the DNA profiles ($6\times 5200 = 31,200$ scan points) makes the use of Shapley values in the kernal SHAP algorithm prohibitively computationally expensive - there are $2^{31200}$ ways to occlude a profile. This also makes DNA profiling a well-suited example for testing a new superpixel framework.  

\section{Results}\label{sec2}

When the CNN was trained, a context window of size 6 × 201 was used around each scan point for classification \cite{taylor_using_2022}. Hence, when investigating the reasoning behind a classification of a point in the EPG only a [6×201] context window around that point requires consideration. In other words, any features outside of this context window are known (or assumed) to not impact the classification. 

Figure \ref{fig:three graphs} shows the results of the focusing occlusion approach. Figure \ref{fig:iter1} shows the profile with the peak being classified indicated by a red box. The ground truth category for the highlighted peak in Figure \ref{fig:iter1} is pull-up, but we consider the Shapley values for classifying it as allelic. We use a top-2 partitioning scheme to demonstrate our method. This means that the top two partitions will be carried through each iteration of the algorithm. The initial focus is on the whole-dye level (Fig \ref{fig:iter2}), and finds that the two most extreme Shapley values correspond to the second and third dye lanes. The Shapley values for each grade from blue (positive influence) to red (negative influence). Dyes without significant Shapley values are not investigated in any further stages. In the second focusing step (which splits the two dye lanes of interest into three segments each) the central regions are deemed most significant to the classification (Fig \ref{fig:iter3}). The final focusing step splits the central areas into three parts each again, and the resulting Shapley values for each are shown in Fig \ref{fig:iter4}. The shade of the colour indicates the magnitude of the Shapley value. The regions of importance identified by the Shapley values align in both strength and direction with the expert-elicited explanation of the classification of the peak as not allelic. Note that while we are talking in terms of the classification of a peak, the method is in fact dealing with the classification of the scan point that is central to that peak (a peak consisting of approximately 10 to 20 scan points). 

By iteratively focusing in on the important blocks, the important features in the input can be analysed without wasting computational power on features with negligible contributions to Shapley values. This reduced the computation from something prohibitively expensive, to analysis of a scan point in seconds. When the top 2 Shapley values are chosen at each step and then split for further processing, Table \ref{tab:timing} shows a full suite of Shapley values for a single scan point can be generated in 1 second. This returns values for classification across all possible categories: Allele, Baseline, Backward Stutter, Forward Stutter, Half Stutter and Pull-up. As such, values for an entire profile can be generated in less than one hour. However, it is unlikely that practical use would require every scan point in a profile to be interrogated. Instead, the likely use case would be to interrogate the assignment of a single peak into a category, leading to a few (or even a single) scan points having Shapley values produced. This would be the same for most time series applications; only a very small subset of all datapoints will be interrogated by an analyst. 

\begin{table}[h!]
    \centering
    \begin{tabular}{p{4cm} p{4cm} |c c c}
       Laptop & Programs & \multicolumn{3}{c}{Run Time (seconds)}  \\
        & & Top 2 & Top 3 & Top 4 \\
       Intel i9-14900HX 2.2GHz 128 GB RAM Nvidia RTX4090 GPU 16GB VRAM & tensorflow V2.10 GPU in R V4.2.3. python V3.10, Cuda V11.8, cuDNN V8.6 & 1 & 3 & 22 \\ 
\hline
    \end{tabular}
    \caption{Computer specifics, programs and run times for generating Shapley values. These results are for a single scan point and generate Shapley values for each of the six possible classifications: baseline, allele, backward stutter, forward stutter, half stutter and pull-up. The results presented are for choosing to focus on the top 2, 3 or 4 Shapley values to guide the splitting step (see Figure \ref{fig:workflow} for a visualisation of the top 2 workflows).}
    \label{tab:timing}
\end{table}

\subsection{Motivating Example}

\begin{figure}[h!]
     \centering
    \begin{subfigure}[b]{0.47\textwidth} 
         \centering
         \includegraphics[width=\textwidth]{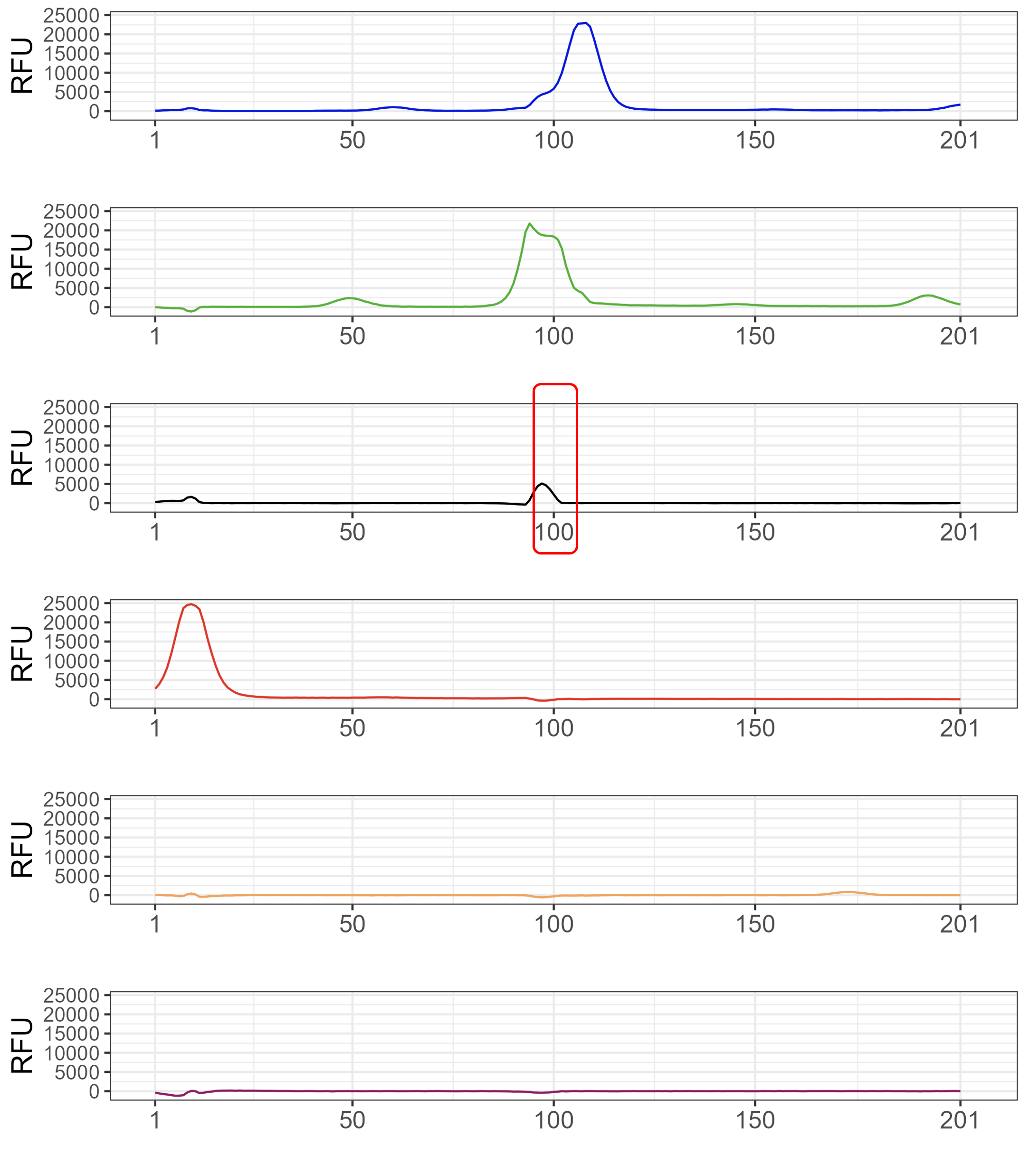}
         \caption{Scan point of interest.}
         \label{fig:iter1}
     \end{subfigure}
     \begin{subfigure}[b]{0.47\textwidth} 
         \centering
         \includegraphics[width=\textwidth]{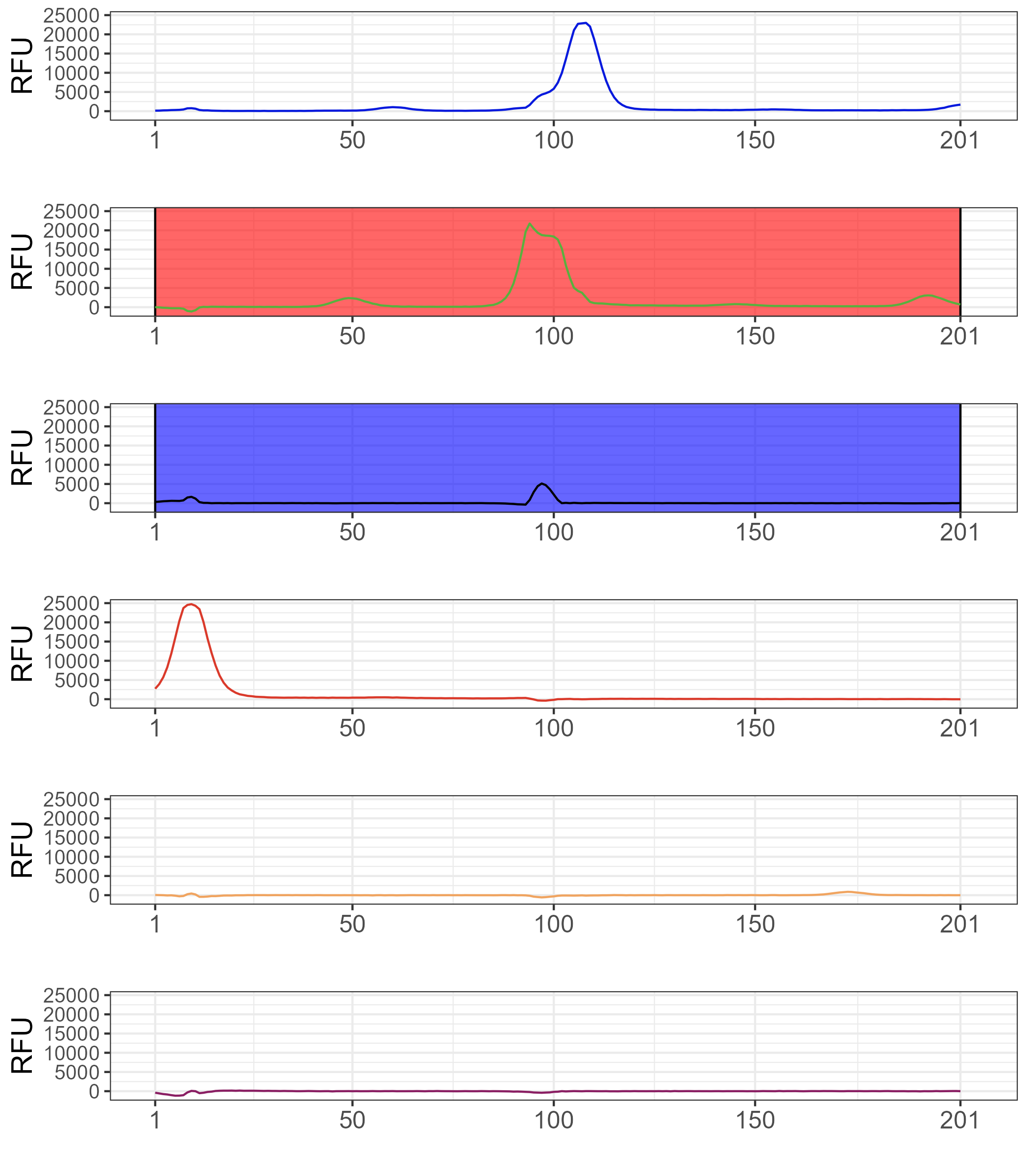}
         \caption{First iteration.}
         \label{fig:iter2}
     \end{subfigure}
     \begin{subfigure}[b]{0.47\textwidth}
         \centering
         \includegraphics[width=\textwidth]{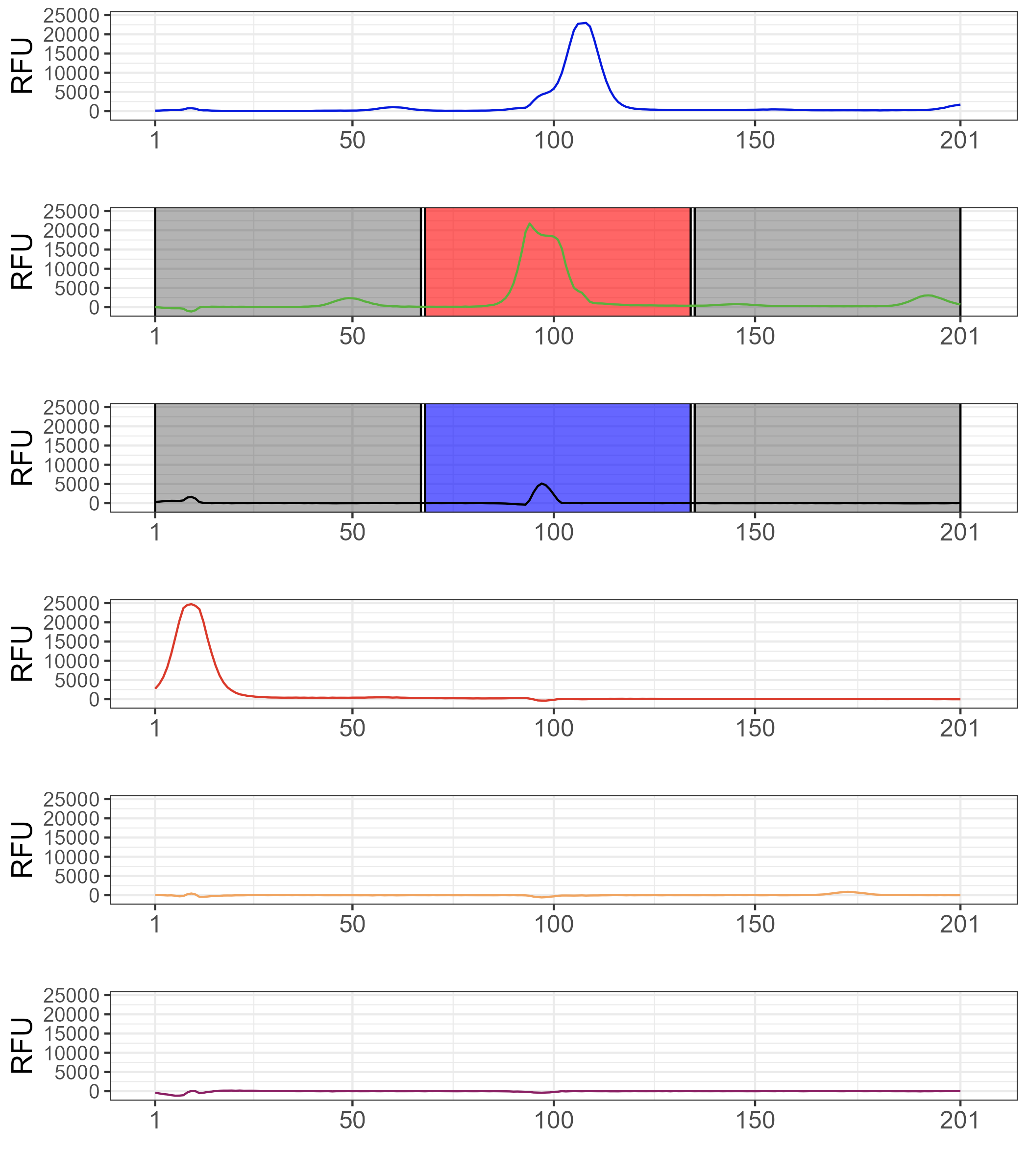}
         \caption{Second iteration.}
         \label{fig:iter3}
     \end{subfigure}
     \begin{subfigure}[b]{0.47\textwidth}
         \centering
         \includegraphics[width=\textwidth]{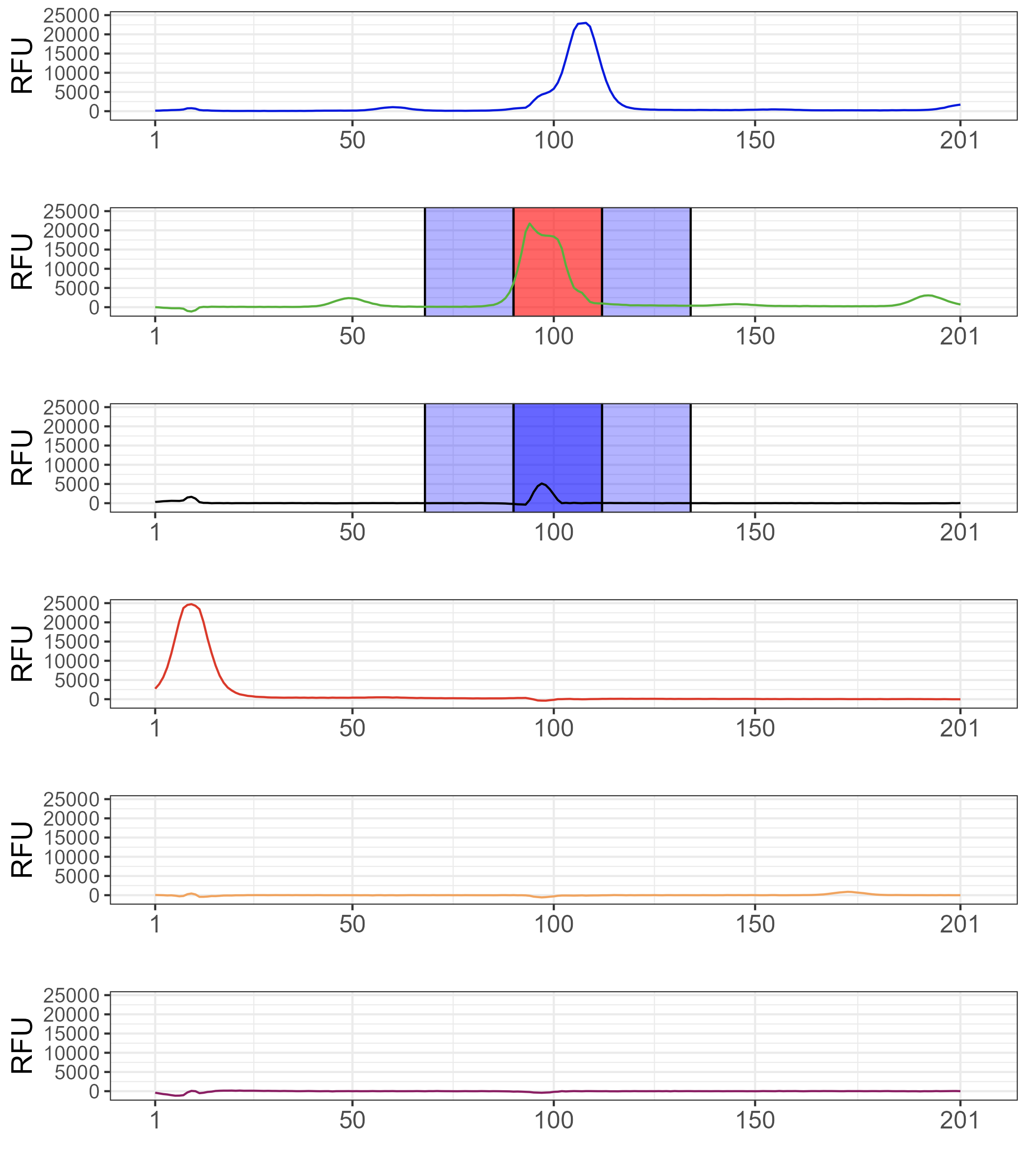}
         \caption{Third iteration.}
         \label{fig:iter4}
     \end{subfigure}
        \caption{An example of isolating features of importance in a DNA profile using the kernel SHAP algorithm, for predicting an allele in dye lane 3. The blue areas correspond to positive Shapley values, the red to negative Shapley values, and the grey areas represent negligible Shapley values. The strength of the colours varies according to the magnitude of the values.}
        \label{fig:three graphs}
\end{figure}

The results in Figure \ref{fig:three graphs} can also be interpreted in the context of the motivating example. Figure \ref{fig:three graphs} shows the results of a typical Shapley value explanation for the classification of the central peak in the third dye lane as allelic. Since the output of the DNA profile classification CNN of \cite{taylor_using_2022} is a vector of probabilities, all Shapley values lie between -1 and +1. A positive Shapley value indicates that the feature is responsible for pushing the prediction toward the classification being made, with a value closer to +1 representing a stronger relationship. A negative Shapley value indicates that the feature is driving the prediction away from the classification.  

Expert elicitation classifies the peak as an artefact type known as a pull-up (the occurrence of a strong signal in one dye lane being detected in another and appearing as a small peak). The small peak in this position provides some support for the region being allelic (as it is definitely not baseline); however, expert elicitation explained that the large peak in the second dye lane suggests that it is a pull-up peak.  We would expect these expert-elicited explanations to be reflected in the results of our XAI approach.

\begin{figure}[h!]
    \centering
    \includegraphics[width = 0.8\textwidth]{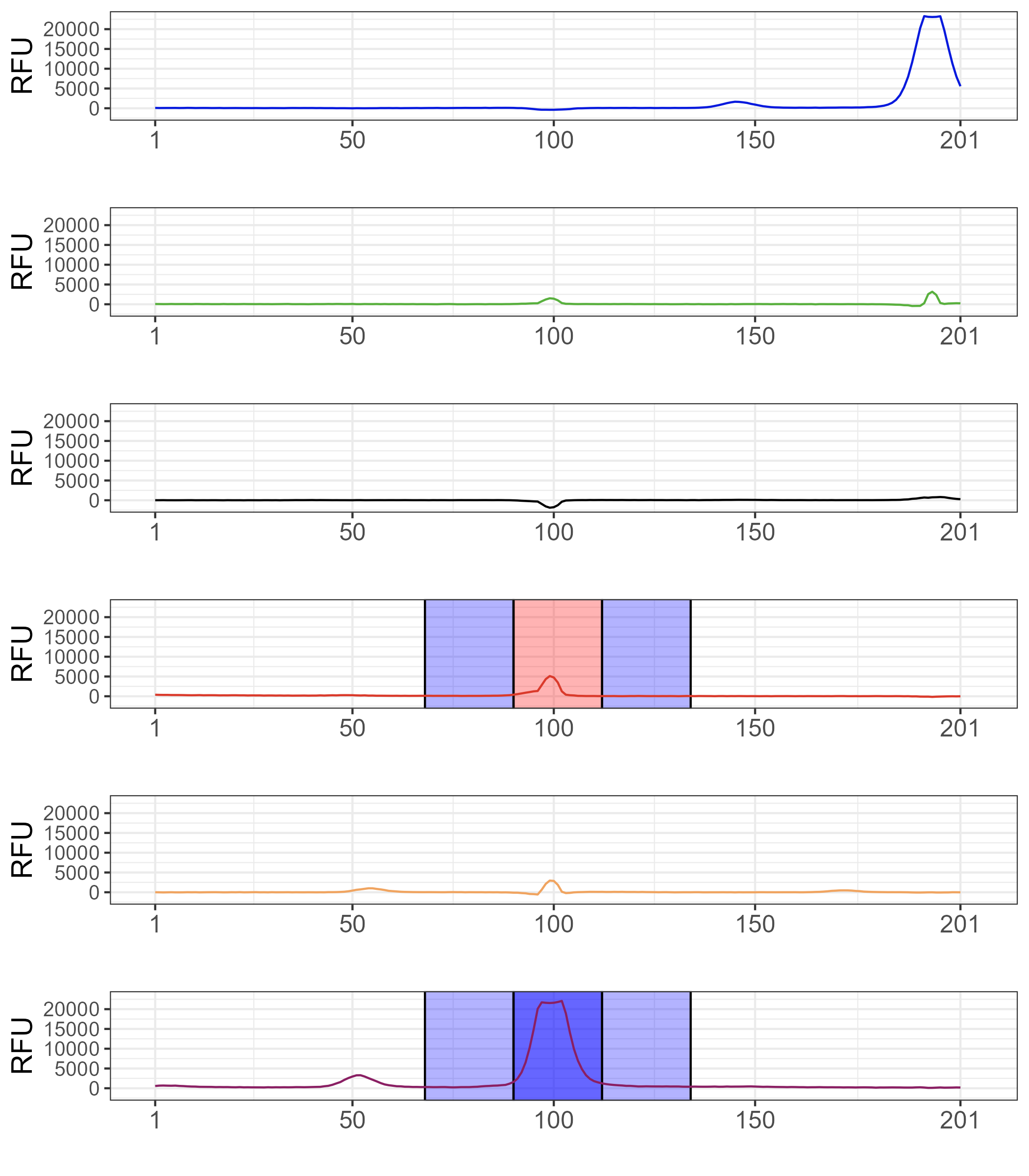}
    \caption{An example of the third iteration of the kernel SHAP algorithm, for classifying a pull-up in the centre of dye lane 4. The blue areas represent positive Shapley values, and the red areas represent negative Shapley values. The remaining areas in the DNA profile had negligible values, indicating that they did not considerably influence the prediction.}
    \label{fig:pull-up-dye4}
\end{figure}

Figure \ref{fig:pull-up-dye4} shows an example of computing the Shapley values for classifying the centre scan point in dye lane 4 as pull-up. Table \ref{tab:shap_vals} displays the Shapley values obtained in the final focusing step. There is a small negative contribution from the block that holds the scan point being interrogated, indicating that the peak features are pushing the prediction away from pull-up and towards an allele. However, there is an overwhelmingly large positive contribution from the centre of dye lane 6 driving the prediction back towards a pull-up. Again, the XAI method shows how the CNN model has learned the patterns that are well known to human readers from their domain expertise. 

\begin{table}[h!]
\centering
\begin{tabular}{ c c c } \hline
Dye lane & Block & Shapley value \\ \hline
4 & 4 & 0.0025 \\
4 & 5 & -0.0286 \\
4 & 6 & 0.0034 \\
6 & 4 & 0.0048 \\
6 & 5 & 0.9359 \\
6 & 6 & 0.0054 \\\hline
\end{tabular}
\caption{The significant Shapley values for predicting a pull-up in the fourth dye lane in the final focussing step of the kernel SHAP algorithm.}
        \label{tab:shap_vals}
\end{table}

\begin{figure}[h!]
    \centering
    \includegraphics[width = \textwidth]{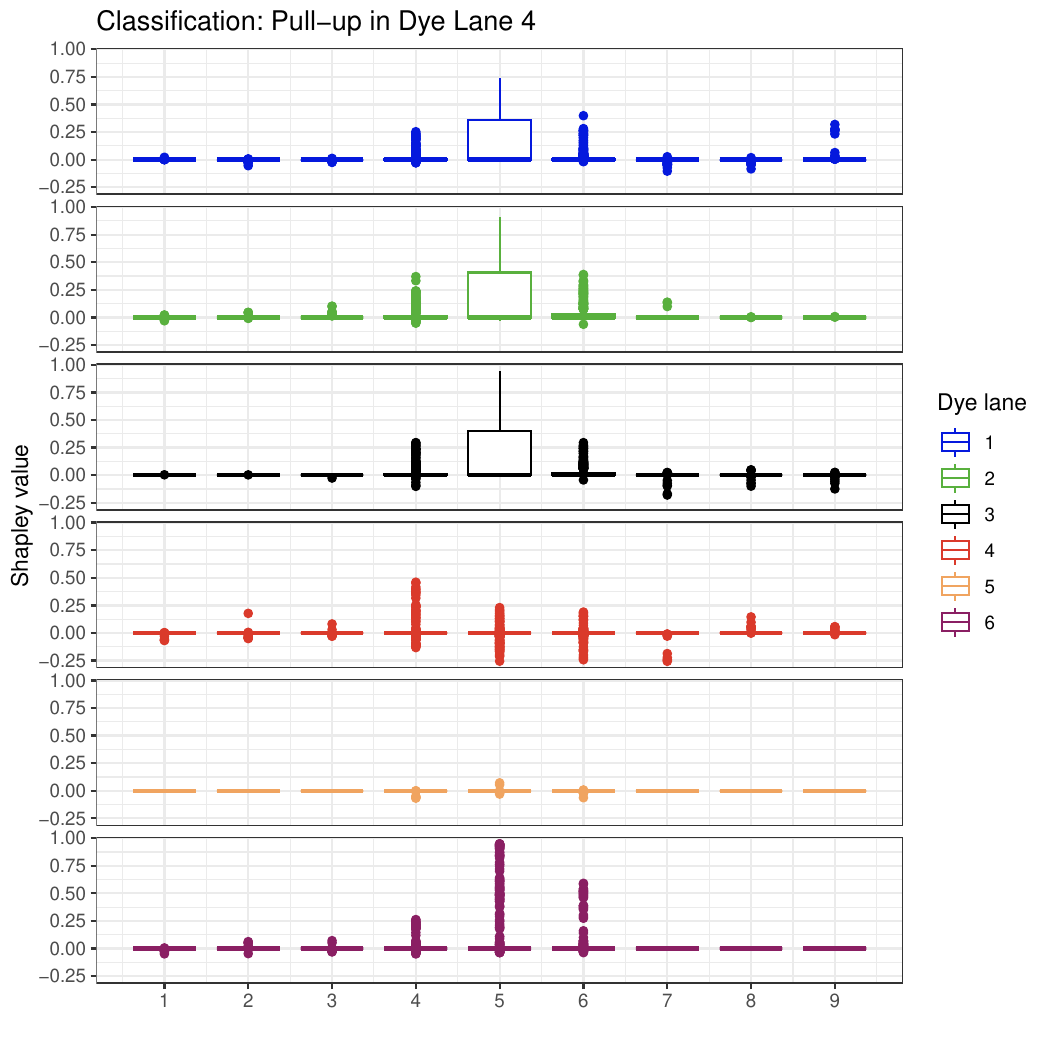}
    \caption{A example of the Kernel SHAP algorithm results for classifying the centre scan point in dye lane 4 as a pull-up, after three focussing iterations. The graph is in the style of a DNA profile, where each row of nine boxplots represents one dye lane, and the x-axis displays the position of each block within its dye lane.}
    \label{fig:general-pullup4}
\end{figure}

Examples thus far have shown the use of the XAI algorithm in the explanation for classification of a single peak in a single profile. Another application of the technique is to consider what regions are important to particular classifications in general. The Shapley values shown in Table \ref{tab:shap_vals} are for the classification of the peak in the fourth dye lane shown in Figure \ref{fig:pull-up-dye4} as being pull-up. By tallying multiple sets of Shapley values for peaks designated as pull-up in dye lane 4, a more holistic view of profile interpretation is achieved (specifically, in this instance, into the important regions in general that lead to classifications of peaks in the fourth dye lane as being pull-up). This could teach scientists about patterns in DNA profiles that they may not have been aware of. Figure \ref{fig:general-pullup4} shows an example of the general results produced from a series of context windows for classifying the centre scan point in dye lane 4 as a pull-up. The graph is in the style of a DNA profile, where each row of nine boxplots represents one dye lane and the x-axis displays the position of each block within its dye lane. There is significant activity in the middle of each of the dye lanes and little contribution in other areas, which is supported by the biology expert elicitation and DNA profiling theory.

Pull-ups are caused by an overlap in the wavelengths being emitted during the creation of the profiles and appear as small peaks in the same scan point location as alleles in other dye lanes. As a result, if an allele is present in another dye lane, it is likely to have caused pull-up in dye lane 4. Hence, there are large positive Shapley values in each of the other dye lanes, depending on where the allele lies. There are also negative values in the central region of dye four as peaks in this area (particularly high-intensity peaks) will drive the classification towards being allelic. In general, the plots created across all dye lanes and classification types reflect expert expectations from their domain expertise, indicating that the model has learned the patterns in the DNA profiles that human readers use in their manual classifications.

\section{Methods}\label{sec11}

\subsection{The data}
The data for this forensic application involves a series of DNA profiles provided by FSSA. Each DNA profile holds 31,200 scan points of interest, corresponding to 5200 scan points in each of the 6 dye lanes. From the original raw profile data of approximately 9000 scan points, the first 2900 scan points and last 900 scan points were discarded because they were outside the range of DNA profiling data leaving [6 x 5200] datapoints per profile. In total, the DNA profile data are stored as a large 3-dimensional tensor of size [$N\times 6 \times 5200$], where `$N$' is the number of profiles in the dataset. 

The aim of reading the DNA profiles is to classify each of these scan points as either an allele, baseline noise, or one of a series of artefacts. The DNA-profile-reading CNN used in this research is currently utilised at FSSA through their one-human-reader and one-model approach. The structure and performance of the CNN are well-defined by Taylor \cite{taylor_using_2022}. The CNN has two initial layers that work on the entire profile, before splitting into six streams for two additional layers, to allow for dye-lane specific features to be learned and classifications to be made in each of the six dye lanes. The CNN can accept a variable number of DNA profiles at once and outputs a vector for each scan point, corresponding to the probabilities of the point being each of the possible features. The final classification is chosen as the feature with the highest probability.

The CNN from \cite{taylor_using_2022} was modified to take inputs of [1 x 6 x 201] scan points but utilising all the same weights and biases, and resulting in outputs of [6 x 8] (6 dye lanes by eight possible fluorescence categories). All computations were carried out using R V4.2.3 \cite{r} with Tensorflow V2.10 GPU \cite{tensorflow}, Keras V2.11.1 \cite{keras}, dplyr V.1.1.4 \cite{dplyr} and SOAR V0.99-11 \cite{SOAR_2013}.

\subsection{Shapley value theory}


A common application in XAI is Shapley regression values. For all subsets $S \subseteq F \backslash \{i\}$, from the set of all features $F$, predictions are made where a feature $i$ is present and withheld. The Shapley value for each feature $i$ is then calculated as defined in Equation \ref{eqn:shap}.

\begin{equation}\label{eqn:shap}
    \phi_i = \sum_{S \subseteq F \backslash \{ i\}} \frac{|S|!(|F|-|S|-1)!}{|F|!} [f(x_{S \cup \{i\}}) - f(x_S)].
\end{equation}

\subsection{Occlusion methods}

Occlusion methods have applications to many types of input data, including images. In this case, the perturbation may involve changing pixels to black, grey or an average of adjacent pixels. In the case of vector or matrix input, zeros or local average values may be used \cite{dardouillet_explainability_nodate}. In
occlusion methods, a predefined window moves across the input and perturbs the data under the window, which is then passed to the model. The occluded features with the greatest variation in the output are considered to have the greatest impact on the decision-making of the model. In the context of Shapley values, groups of occlusion blocks are referred to as coalitions. In this research, blocks are occluded by replacing the values in the profile with a vector of zeros.

However, as the number of features increases, the number of possible combinations (coalitions) increases exponentially, making the method computationally expensive. One solution typically used to manage this is to subsample the coalitions $S$ rather than evaluating all $2^F$ combinations. However, using the partitioning method, we describe all exhaustive sets of coalitions used in the calculation of Shapley values.

\subsection{Kernel SHAP}

In 2017, Lundberg and Lee proposed a Shapley value application that combines components of LIME and SHAP, named Kernel SHAP. Kernel SHAP is more efficient in sampling Shapley values, and works by transforming Equation \ref{eqn:shap} into a weighted least squares problem \cite{chau_rkhs-shap_2022}. In their research, it was found that by appropriately weighting the regression, Shapley values are recovered as the model coefficients \cite{lundberg_unified_2017}. Using data $x$, model $f$, and a subset of features $S \subseteq F$, a linear model $g(z') = \phi_0 + \sum_{i=1}^M \phi_i z’_i$ is fit, where $z' \in \{0,1\}^M$ is a binary vector of length $|F|$, denoting whether or not features are included in $S$. The loss function $L$ and corresponding weighting kernel $\pi_z$ are shown in Equation \ref{eqn:kernelshap}.

\begin{equation}\label{eqn:kernelshap}
\begin{split}
L(f, g, \pi_z) &= \sum_{z \in Z} [f(h(z)) - g(z) ]^2 \pi_z (z), \\
\text{where } \pi_z &= \frac{(M-1)}{{M \choose |z|} |z| (M - |z|)}, \\
\text{ and } h(z) \text{ maps } &z \text{ to the occluded input, which is fed into the model } f(.).
\end{split}
\end{equation}

Solving for the $\pmb{\phi}$ coefficient vector gives
\begin{equation}
\begin{split}
\pmb{\phi} &= (Z^T W Z)^{-1} Z^T W Y, \\
&\text{where } Z  \text{ is a matrix with rows } z_i ,\\
&W \text{ is the diagonal matrix with } W_{ii} = \pi_z (z_i), \\
&\text{and } Y \text{ is the vector of predictions } f(h(z_i)).
\end{split}
\end{equation}

\subsection{Focusing the Shapley value algorithm}

In the first step, each context window of [6 x 201] datapoints is split into individual [1 x 201 ] dye lanes, which are treated as blocks to be occluded. The kernel SHAP algorithm is then applied to the six blocks (using all 64 possible occlusion combinations), and Shapley values are calculated using the kernel SHAP method. The two blocks with the most extreme Shapley values (those the farthest from zero) are marked for inclusion in the next round of focusing. 

In focus round 2 the two previously identified blocks (dye lanes) are split into three sections for further investigation. Again, all combinations of occlusions in the 6 blocks (coming from the two dye lanes each split into three sections) are used with the kernel SHAP algorithm. The dye lanes that were not marked in focus round 1 are not changed from their original values. Again, the two blocks (one third of a dye lane) that hold the most extreme Shapley values are marked for use in the third and final focus round.

In the final focus round, the two blocks marked in the second round are again split into three sections and these six new blocks have all occlusion combinations generated and Shapley values calculated using the kernel SHAP algorithm. As there are no further focusing rounds, all six Shapley values are then overlaid on the original DNA profile to designate where the importance regions lie. This process is run for each fluorescence category for a scan point of interest. After the final round of the process the blocks of scan points (i.e. the equivalent structures to superpixels for image analysis) are approximately 23 scan points, which is approximately the span of a peak.

\begin{figure}[h!]
    \centering
    \includegraphics[width = \textwidth]{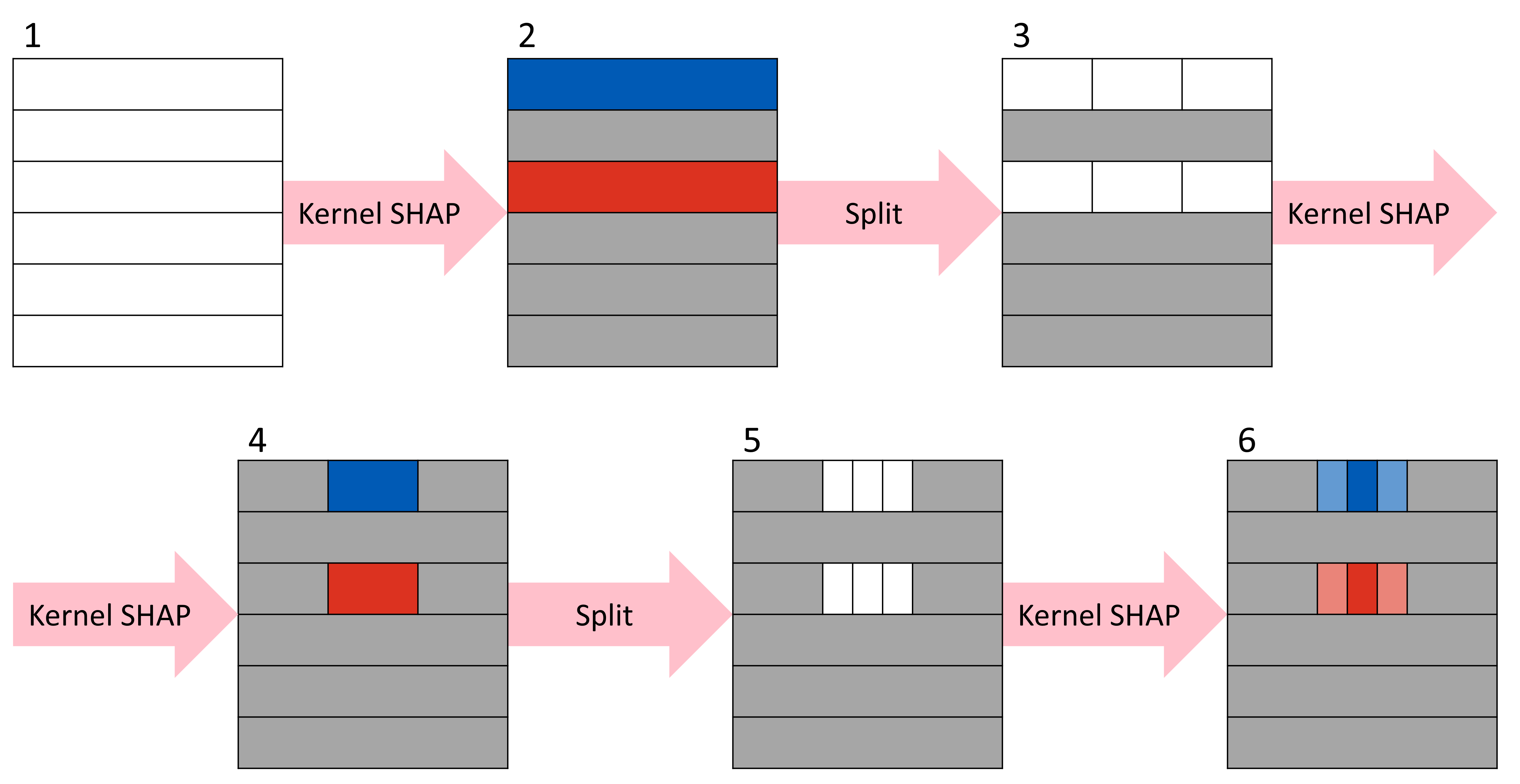}
    \caption{The workflow of the kernel SHAP iteration process. In the first step, the \textit{kernal SHAP} algorithm is applied to each of the six dye lanes, highlighting the top two Shapley values which are either positive (\textit{blue}) or negative (\textit{red}). The top two selected regions are then \textit{split} into three regions, and the process continues until the desired granularity is reached.}
    \label{fig:workflow}
\end{figure}

Figure \ref{fig:workflow} shows this iterative process for one scan point being interrogated for one fluorescence category. In Figure \ref{fig:workflow}, red designates an area that negatively influences the classification, blue represents an area that positively influences a classification, white represents an area that is yet to be interrogated and grey represents areas that have been interrogated and found to have negligible influence on classification.

\subsection{Shapley values for multiple profiles}

To investigate the areas of DNA profiles that were important for classifications in general (and not for just one scan point), the focusing Shapley method was applied to all nonbaseline scan points across 19 complete DNA profiles. All Shapley values for combinations of fluorescence category and dye lane were grouped (e.g. allele in dye 1, pull-up in dye 1, ……, pull-up in dye 6). For each combination of dye lane and fluorescence category, the distribution of the Shapley values were displayed in box plots corresponding to the regions in the final focusing layer.

\section{Discussion}\label{sec12}

There are various use cases for the specific and general approaches for the Shapley values and plots created. Examining the specific Shapley values allows a scientist to isolate and interrogate the classification of a single peak in a profile, to determine what influenced the CNN in making its decision. In doing so, they gain a deeper understanding of the operation of the CNN, which can be used to provide information to stakeholders (such as the courts) on why the system has classified peaks in the way it has.  

The general case allows for a more holistic view of profile interpretation, for example, finding which areas are important in classifying a point as an allele. This has the potential advantage of teaching scientists about patterns in DNA profiles that they may not have been aware of.

Currently at FSSA, a one-human-reader and one-CNN-model approach is adopted. This research provides a deeper understanding of how the model reads the profiles, such that the sole use of CNN analysis of DNA profiles in criminal investigations can be justified in a court of law. It can also be used as part of a validation process to ensure the proper functioning of a CNN. If there are questions about a classification from the model, or if experts are called upon to justify any classifications, the kernel SHAP algorithm can be used to determine which areas of the DNA profile have the greatest predictive impact. As a result, the way in which the CNN reads the DNA profiles is more transparent and explainable.

There are several directions to build on this research. First, the effect of the block size could be investigated. Here, the final block size after the final focusing step was 22-23 scan points long, with nine blocks in each row. Testing the effect could include further iterations of focusing, or increasing the number of blocks into which each iteration splits. Taking more than the most extreme two Shapley values in each iteration could be replaced with a threshold approach to allow for a variable number of features to be investigated in each iteration of the kernel SHAP algorithm and for each DNA profile. Although some preliminary results of this kind are displayed in Table \ref{tab:timing}, understanding the value of widening the number of considered Shapley values needs further consideration. The current methodology (with regard to each of these factors listed) is fit for purpose in the forensic application and limits the computational load to $3 \times 2^6 = 192$ Shapley value calculations, which were completed in only seconds of computation time.  

When occluding blocks in the profiles, the values were replaced with a vector of zeros. An extension of this could include the use of baseline noise for the occlusions rather than zeros to have the appearance of areas of DNA baseline, rather than values that the CNN model has not seen before (i.e. a string of zeros). The use of zeros, however, did not seem to affect the performance of the CNN model or the XAI model to any significant level in this application. Alternatively, a counterfactual approach could be taken, where the blocks are replaced with additional peaks in the profile rather than occluding. The aim of this work is to include the kernel SHAP method described here into an existing  user-friendly tool (FaSTR™ DNA \cite{lin_developmental_2021}) for investigating peak classifications. 

\section{Conclusion}

This research has developed an efficient, novel application of Shapley values and the kernel SHAP algorithm that has the potential to widen the scope of applications for this powerful method. In doing so, it has provided greater insights and transparency for DNA evidence in criminal investigations, while simultaneously reducing the workload of forensic experts and increasing the potential accuracy of DNA analysis on the whole.

The potential use cases of this approach is not limited to DNA analysis. This research has demonstrated that Shapley values can provide insight for vectorised, numeric, time-series-like data. In any field where data of this type are used, and where transparency and accountability are desired, this contributes to the pathway forward. Fields such as finance and business may find great use in explainable models, such as these.

\newpage
\backmatter

\bmhead{Acknowledgments}
Not applicable

\section*{Declarations}

Not applicable

\end{document}